\documentclass[pre,floatfix,twocolumn,showpacs]{revtex4}
\usepackage{epsfig}
\usepackage{amsmath}
\usepackage{graphicx}
\usepackage{dcolumn}
\usepackage{bm}
\usepackage{amssymb}
\usepackage{amsmath}
\usepackage{epsf}
\usepackage{subfigure}
\usepackage{epstopdf}
\usepackage{color}
\usepackage{subeqnarray}
\usepackage{mathrsfs}
\usepackage[colorlinks=true, pdfstartview=FitV, linkcolor=red, citecolor=blue, urlcolor=blue]{hyperref}
\newcommand{\be}{\begin{equation}}
\newcommand{\ee}{\end{equation}}
\newcommand{\ben}{\begin{eqnarray}}
\newcommand{\een}{\end{eqnarray}}
\newcommand{\bra}[1]{\langle #1|}
\newcommand{\ket}[1]{|#1\rangle}

\newcommand{\dg}{\dagger}
\newcommand{\bsen}{\begin{subeqnarray}}
\newcommand{\esen}{\end{subeqnarray}}

\begin{document}
\title{Approach to Equilibrium of a Nondegenerate Quantum System: Decay of Oscillations and 
Detailed Balance as Separate Effects of a Reservoir}
\author{M. Tiwari}\email{mukesh\_tiwari@daiict.ac.in}
\affiliation{D A Institute of Information and Communication Technology (DA-IICT), Gandhinagar-382007, India}
\author{V. M. Kenkre}\email{kenkre@unm.edu}
\affiliation{Consortium of the Americas for Interdisciplinary Science and Department of Physics and Astronomy\\
University of New Mexico, Albuquerque, NM 87131, USA}
\begin{abstract}
The approach to equilibrium of a nondegenerate quantum system involves the damping of microscopic 
population oscillations, and, additionally, the bringing about of detailed balance, i.e. the 
achievement of the correct Boltzmann factors relating the populations. These two are separate 
effects of interaction with a reservoir. One stems from the randomization of phases and the other 
from phase space considerations. Even the meaning of the word `phase' differs drastically in the 
two instances in which it appears in the previous statement. In the first case it normally refers to 
quantum phases whereas in the second it describes the multiplicity of reservoir states that 
corresponds to each system state. The generalized master equation theory for the time evolution of 
such systems is here developed in a transparent manner and both effects of reservoir interactions are
addressed in a unified fashion. The formalism is illustrated in simple cases including in the 
standard spin-boson situation wherein a quantum dimer is in interaction with a bath consisting of 
harmonic oscillators. The theory has been constructed for application in energy transfer in molecular aggregates and in photosynthetic reaction centers.
\end{abstract}
\maketitle
\section{Introduction}
\begin{figure}[!h]
\centering
\includegraphics[viewport = 10 0 300 200, width = 70mm]{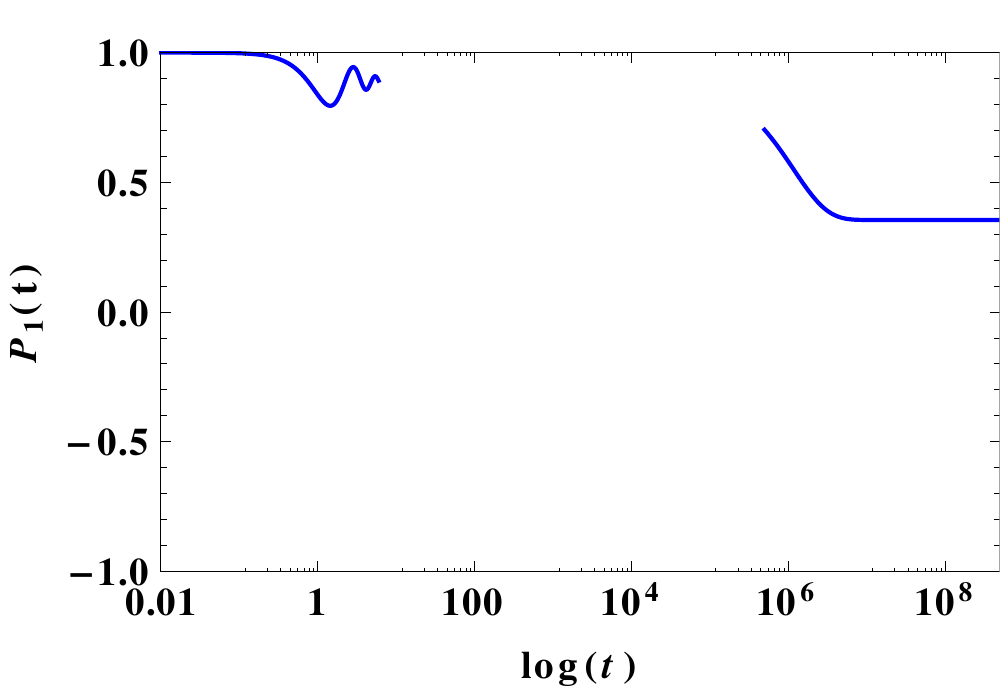}
\caption{Partial trace of the short and long time behavior of the time evolution of the initially occupied site $P_{1}$ for a nondegenerate dimer.}\label{fig:partialpl}
\end{figure}

The purpose of this paper is to develop a generalized master equation (GME) formalism for the 
description of quantum evolution of a system in interaction with a reservoir when the system is 
\emph{nondegenerate} in that the energies of the system are different from one another. 
The situation is ubiquitous in chemical physics, an important real situation of interest 
being the transfer of electronic excitation in a pair of unlike molecules such as tetracene and 
antharacene. Another example of interest, relevant to biophysics, is the capture of energy by a 
photosynthetic reaction center from a molecule of the antenna. Energy transfer in photosynthesis 
has had a venerable history \cite{{history1},{history2},{history3},{history4},{history5}} decades ago and is coming under active 
investigation \cite{{Scholes2010},{Fleming2009}} in recent times as well.
The case of a \emph{degenerate} system is relatively easy to understand or describe because 
the interactions of the system with a reservoir such as a bath of phonons leads only to 
randomization and consequent loss of quantum coherence, and a transition from wavelike motion or 
reversible oscillations of probability differences to incoherent motion or irreversible decays. 
The probability difference $p(t) = P_{1}(t)- P_{2}(t)$ between the two localized states $1$ and $2$, 
depicting for instance the difference in the occupation of two molecules of interest by the 
excitation, is naturally described by
\be
\frac{dp(t)}{dt} + 2\int_{0}^{t}\mathcal{W}(t-s) p(s) ds = 0.
\ee
The time dependence of $\mathcal{W}(t)$ determines, in appropriate fashion, the transition from coherent to incoherent behavior. Bath interactions decide the decay behavior of $\mathcal{W}(t)$ which, in turn, dictates the loss of quantum coherence. However, in the nondegenerate system under consideration in the present paper, additional questions arise from the fact that two unequal memories exist, $\mathcal{W}_{12}$ and $\mathcal{W}_{21}$, one for each direction of transfer. Is detailed balance obeyed by these two memories the way it is by their time integrals, i.e., by the rates $F_{12} = \int_{0}^{t}dt~\mathcal{W}_{12}(t)$ and $F_{21} = \int_{0}^{t}dt~\mathcal{W}_{21}(t)$? Does this mean that $\mathcal{W}_{12}(t)$ and $\mathcal{W}_{21}(t)$ are in the Boltzmann ratio at every instant of time? Does such a time-independent ratio relation provide an accurate description of transfer at all times? In other words, is it possible to have a separation situation of the memory in the form $\mathcal{W}_{12}(t) = F_{
12} \phi(t)$? If so, how would
\be\label{eq:dimerf}
\frac{dp}{dt} + (F_{12} + F_{21}) \int_{0}^{t} \phi(t-s) p(s) ds = (F_{12} - F_{21})\int_{0}^{t}\phi(s)ds
\ee
provide an accurate description of both the randomization process (which would be taken care of by the time dependence of $\phi(t)$ as in the degenerate case) and the detailed balance process which would depend on the energy state difference of the system which are \emph{independent} of the reservoir?
We provide explicit answers to these questions below. They can be put to practical use in the accurate description of short time and long time evolution of excitation transfer in dissimilar molecule pairs as encountered in the study of molecular aggregates and photosynthetic systems. The basis for the solution we provide is a detailed formulation in terms of projection operators, implementation where necessary from generalizations of the spectral prescription of F\"{o}rster \cite{Forster1948}, and computations from the resultant memory all of which can be found in a paper written almost 40 years ago\cite{Kenkre1974}. Other approaches \cite{{Silbey1971},{Silbey1975}, {Rackovsky1973}} could be employed for similar questions but in the light of relations that are available \cite{{Kenkrer1},{Kenkre1975}} between various formalisms, we will use here only the GME approach.
Our purpose in providing the explicit analysis below is to clarify confusing issues that our perusal of literature has uncovered. The need for such an analysis might be appreciated by noticing that an equation such as (\ref{eq:dimerf}) would not be able to describe correctly the time evolution expected at both short and long times as displayed in Fig.~\ref{fig:partialpl}. In that Figure we show a partial plot of the probability $P_{1}(t)$ of the initially occupied site in the quantum nondegenerate dimer. Time is measured in units of $2 V$, arbitrary values of coupling to the bath, temperature and field being assumed for illustrative purposes. Two distinctly different behaviors can be immediately identified, oscillations at short times, and approach to a final steady state value. The details of both are influenced by the energy difference $2\Delta$ between the two states, $1$ and $2$, in a way not present in the evolution of a degenerate dimer. The long time behavior is determined by the exponential factor 
$exp(-2\Delta/k_BT)$, where $T$ is the temperature, in that it equals the ratio of the steady state values of the two probabilities. The short time behavior has nothing to do with temperature and exhibits the specific effects of the lack of resonance of the two states. Unlike the long time behavior that is certainly sensitive to the sign of $\Delta$ (for instance to which of the two is the more energetic molecule), the short time behavior is independent of whether the energetically lower state or higher state is initially occupied. Specifically, for
site 1 being initially occupied, $P_{1}(t)$ is
\be
P_{1}(t)= \cos\left(\Omega t\right) + \frac{1 + 2\Delta^{2} }{2\Omega^{2}}\left(1-\cos\left(2 \Omega t\right)\right)
\ee
with $\Omega=\sqrt{\Delta^{2} + 1}$, and $\Delta$ is normalized with $V$. Here, and everywhere in this paper, we set $\hbar=1$ for simplicity in notation.
The starting point for our considerations is the Liouville equation
\be\label{eq:liouville}
i\frac{\partial \rho}{\partial t} = \left[H,\rho\right] = \mathcal{L}\rho
\ee
where $\rho$ is the total density matrix of the system, $\mathcal{L}$ is the Liouville operator and $H$ is the total Hamiltonian of the system. The Hamiltonian is generally expressed in terms of the unperturbed part $H_{0}$ and a perturbation $V~(H=H_{0} + V)$ and the Liouville operator is similarly decomposed as $\mathcal{L} = \mathcal{L}_{0} + \mathcal{L}_{v}$.
The endpoint after the use of appropriately chosen projection operators that generalize the Zwanzig's diagonalization\cite{Zwanzig1961} with coarse graining over the reservoir is\cite{Kenkre1974}
\be\label{eq:gme}
\frac{\partial P_{M}}{\partial t} = \int_{0}^{t}ds~\sum_{N}\mathcal{W}_{MN}(t-s)P_{N}(s) - \mathcal{W}_{NM}(t-s)P_{M}(s).
\ee
Here $M,~N$ denote system states such as $1$ and $2$ signifying occupation by the excitation of the molecules, details of the reservoir and the interaction of the system with it have all gone into the formation of the detailed form of the memories $\mathcal{W}(t)$, and the initial random phase approximation has been employed.
As expected, in the following sections we will see that our analysis will yield for a nondegenerate quantum dimer, not Eq.~(\ref{eq:dimerf}), but
\ben\nonumber\label{eq:2sitegme}
\frac{dp(t)}{dt} &+ \int_{0}^{t}~ds \left[\mathcal{W}_{12}(t-s) + \mathcal{W}_{21}(t-s)\right]p(s)&\\
&= \int_{0}^{t}~ds \left[\mathcal{W}_{12}(s) - \mathcal{W}_{21}(s)\right],
\een
and, importantly, in that the time dependence of the memories $\mathcal{W}(t)$ will naturally take care of \emph{both} effects of the reservoir, randomization with loss of coherence and Boltzmannization with proper thermal behavior.
The paper is laid out as follows. In Section \ref{sec:gmereview} we present the GME formalism suitable to nondegenerate systems, discussion of the salient features of the time dependent memories and the connection between short and long time dynamics is presented in Section \ref{sec:memprop}. In Section \ref{sec:nddimer} we apply this formalism to the standard model of a nondegenerate quantum dimer, while conclusions are presented at the end. 
\section{Form of the memories for a nondegenerate quantum dimer}\label{sec:gmereview}
The derivation of the GME proceeds with the application of appropriately chosen projection operators on Eq.~(\ref{eq:liouville}), which, in conjunction with the weak coupling approximation and the initial diagonalization assumption, results in
\be\label{eq:gmeint}
\frac{\partial \mathcal{P}\rho}{\partial t} = -\int_{0}^{t} ds \mathcal{P}\mathcal{L}_{v}G_{0}(t-s)\mathcal{L}_{v}\mathcal{P}\rho(s)
\ee
for the relevant part of the density matrix. The simplified form $G_{0}(t) = \exp(-it \mathcal{L}_{0})$ on the right in Eq.~(\ref{eq:gmeint}) is a consequence of the weak-coupling approximation. We specifically use the coarse graining projection operator\cite{Kenkre1975} defined as
\be\label{eq:projection}
\bra{M,m}\mathcal{P}\mathcal{O}\ket{N,n} = \frac{e^{-\beta \mathcal{E}_{m}}}{\mathcal{Q}}\sum_{m'}\bra{M,m'}\mathcal{O}\ket{M,m'}\delta_{M,N}\delta_{m,n}
\ee
whose operation on any operator $\mathcal{O}$ has three consequences. It diagonalizes the operator $\mathcal{O}$ in the eigenstates of $H_{0}$ as shown by the Kronecker deltas in M,N and m,n; it traces over the bath as shown by the summation over m; and it thermalizes with the phonon equilibrium density matrix as shown by the Boltzmann factors $\exp(-\beta \mathcal{E}_{m})$. In Eq.~(\ref{eq:projection}), $\beta = 1/k_{B}T$, $\mathcal{E}_{m}$ and $\mathcal{E}_{n}$ are the energies of the bath states over which coarse-graining is performed, $M, N$ are molecule states and $\mathcal{Q} = \sum_{m}e^{-\beta\mathcal{E}_{m}}$ is the normalization factor of the bath equilibrium density matrix. As is well known, the definition of $\mathcal{P}$ in Eq.~(\ref{eq:projection}) and Eq.~(\ref{eq:gmeint}) results in the generalized master equation (Eq.~(\ref{eq:gme})) for the probability of occupancy of site $M$ given by
\be
P_{M} = \sum_{m}\bra{M,m}\rho\ket{M,m}.
\ee
Also well known is the form of the memories appearing on the right in Eq.~(\ref{eq:gme}):
\bsen
\label{eq:memory}
\slabel{eq:memory12}\nonumber
\mathcal{W}_{MN}(t)&=& \frac{2}{\mathcal{Q}}\sum_{m,n} e^{-\beta\mathcal{E}_{n}}\left|\bra{M,m}V\ket{N,n}\right|^{2}\\
&& \times\cos\left\{\left(\mathcal{E}_{mn}+\mathcal{E}_{MN} \right)t\right\},~\rm{and}\\\nonumber\slabel{eq:memory21}
\mathcal{W}_{NM}(t)&=& \frac{2}{\mathcal{Q}}\sum_{m,n} e^{-\beta\mathcal{E}_{m}}\left|\bra{M,m}V\ket{N,n}\right|^{2}\\
&&\times \cos\left\{\left(\mathcal{E}_{mn}+\mathcal{E}_{MN} \right)t\right\}
\esen
and involves a thermal average over the initial and sum over the final vibrational states. $\mathcal{E}_{MN} = \mathcal{E}_{M} - \mathcal{E}_{N}$ and $\mathcal{E}_{mn} = \mathcal{E}_{m} - \mathcal{E}_{n}$ are the energy differences between the system and bath states respectively. Note that the only difference to be discerned in the two expressions (a) and (b) representing transfer in opposite directions between the system states $M$ and $N$ is the appearance of $\mathcal{E}_{n}$ in (a) but $\mathcal{E}_{m}$ in (b). This is in keeping with the textbook statement \cite{Zwanzig2001} that there should be an average for initial states and sum over final states in such an expression. By changing bath indices $n$ and $m$ within the summation, making the physical assumption that the system and bath are independent entities, and taking advantage of the insensitivity of the cosine to the sign of its argument, we can rewrite the second memory above as
\bsen
\label{eq:memory10b}
\nonumber
\mathcal{W}_{NM}(t)&=& \frac{2}{\mathcal{Q}}\sum_{m,n} e^{-\beta\mathcal{E}_{n}}\left|\bra{M,m}V\ket{N,n}\right|^{2}\\
&&\times \cos\left\{\left(\mathcal{E}_{mn}-\mathcal{E}_{MN} \right)t\right\}.\nonumber
\esen
This form of (10b) differs from (10a) only in the sign of the system energy difference $\mathcal{E}_{MN}$. We will explore below the important question concerning what features of the bath, if present, would lead to proper thermalization of the system probabilities.
Important to emphasize is that these expressions are not exact but arise from the weak-coupling approximation that cannot be avoided in a practical calculation. How much of physically expected behavior of the system is retained by their approximated form is not clear \emph{a priori}. Eqs.~(\ref{eq:memory}) in their current form do not make manifest the spectral features of the bath. To clarify these features, we convert Eqs.~(\ref{eq:memory}) into their integral form (see Appendix of Ref.~\cite{Kenkre1977}) by introducing the density of bath energy states and the continuous variable $z = \mathcal{E}_{m}-\mathcal{E}_{n}$, and bundling a product of various quantities under the symbol $\mathcal{Y}(z)$:
\bsen\label{eq:ratesintform}
\slabel{eq:ratesintform1}
\mathcal{W}_{MN}(t)&=& \int_{-\infty}^{\infty} dz~\mathcal{Y}(z)\cos\left[\left(z+\mathcal{E}_{MN}\right)t\right]\\\slabel{eq:ratesintform2}
\mathcal{W}_{NM}(t)&=&\int_{-\infty}^{\infty} dze^{-\beta z} \mathcal{Y}(z)\cos\left[\left(z+\mathcal{E}_{MN}\right)t\right].
\esen
It is instructive once again to rewrite (\ref{eq:ratesintform2}) in the alternate form
\be\label{eq:wnmmodified}
\mathcal{W}_{NM}(t)=\int_{-\infty}^{\infty} dz~\mathcal{Y}(z)\cos\left[\left(z-\mathcal{E}_{MN}\right)t\right].
\ee
The passage from (\ref{eq:memory}) to (\ref{eq:ratesintform}) is made by splitting the bath summation in the former first into a primed summation over bath states which have the energy difference $z = \mathcal{E}_{m}-\mathcal{E}_{n}$, the multiplication by the number of bath states $dz\rho(z)$ having the given value of the energy difference $z$ and a subsequent summation (integration) over $z$ that appears in (12). Thus, in these expressions $\mathcal{Y}(z)$ is itself given by
\be\label{eq:yzintform}
\mathcal{Y}(z) = \frac{2\rho(z)}{\mathcal{Q}}\sum_{m,n}^{\prime} e^{-\beta\mathcal{E}_{n}}\left|\bra{M,m}V\ket{N,n}\right|^{2}
\ee
where the prime over the summation restricts it as explained above. The function $\mathcal{Y}(z)$ given by Eq.~(\ref{eq:yzintform}) contains all the necessary information regarding the transitions by taking into account the spectral features of the bath and henceforth will be referred to as the spectral function as elsewhere in the literature. The time evolution of the probability difference in a quantum dimer can be obtained from Eq.~(\ref{eq:2sitegme}) after substituting the appropriate memories $\mathcal{W}_{12}$ and $\mathcal{W}_{21}$ which are most generally (but within the weak coupling, i.e., perturbative, approximation) given by Eqs.~(\ref{eq:ratesintform}) and are determined from the interaction and a knowledge of the density of states.

\section{General memory description of detailed balance and decay of oscillations}\label{sec:memprop}

Having obtained the basic form for the time dependent memory function in the previous section, we now examine the features of the spectral function which ultimately determines the behavior of the memories. For a system that approaches equilibrium the long time limit of the memories are given by their Fermi rates counterpart $\left(F= \int_{0}^{\infty} dt' \mathcal{W}(t')\right)$. In the case of a two-site nondegenerate system with energy difference $\mathcal{E}_{12}=2\Delta$ between the two sites the ratio of the rates in the long time limit is given by the detailed balance condition,
 
\be\label{eq:detailedbal1}
\frac{F_{12}}{F_{21}} = \frac{\int_{-\infty}^{\infty}dz~\mathcal{Y}(z)\delta(2\Delta + z)}{\int_{-\infty}^{\infty}dz~e^{-\beta z}\mathcal{Y}(z)\delta(2\Delta + z)} = e^{-2\beta\Delta}\cdot
\ee
In addition, using Eq.~(\ref{eq:wnmmodified}) for $\mathcal{W}_{21}$, the ratio of the two rates in the long time limit can also be written as, 

\be\label{eq:detailedbal2}
\frac{\int_{-\infty}^{\infty} dz \mathcal{Y}(z)\delta(2\Delta + z)}{\int_{-\infty}^{\infty} dz \mathcal{Y}(z)\delta(2\Delta - z)}= \frac{\mathcal{Y}(-2\Delta)}{\mathcal{Y}(2\Delta)}.
\ee
Eqs.~(\ref{eq:detailedbal1}) and (\ref{eq:detailedbal2}) gives $\mathcal{Y}(-z) = \mathcal{Y}(z)e^{-\beta z}$ implying that $\mathcal{Y}(z)e^{-\beta z}$ is the mirror image of $\mathcal{Y}(z)$. While calculating the time dependent memories in the GME it is therefore sufficient to obtain only the spectral function $\mathcal{Y}(z)$. These memories, when incorporated into the probability equations, give the dynamics for the specific system of interest. To understand the resulting dynamics it is instructive to examine Eq.~(\ref{eq:2sitegme}) in the Laplace domain. For the population initially at site $1$ the Laplace transform of Eq.~(\ref{eq:2sitegme}) gives 

\be\label{eq:plap}
\tilde{p}(\epsilon) = \frac{1}{\epsilon}\tilde{\xi}(\epsilon) + \frac{1}{\epsilon + \tilde{\mathcal{W}}_{12}(\epsilon) + \tilde{\mathcal{W}}_{21}(\epsilon)}\left[1 - \tilde{\xi}(\epsilon)\right]
\ee
with $\tilde{\xi}(\epsilon) = \frac{\tilde{\mathcal{W}}_{12}(\epsilon) - \tilde{\mathcal{W}}_{21}(\epsilon)}{\tilde{\mathcal{W}}_{12}(\epsilon) + \tilde{\mathcal{W}}_{21}(\epsilon)}$ as the ratio of the difference and the sum of the Laplace transform of two memories, $\epsilon$ being the Laplace variable. The value of the probability difference at long times from the Abelian theorem, which equates $\lim_{\epsilon\rightarrow 0} \epsilon\tilde{p}(\epsilon)$, is
\be\label{eq:lontimelap}
p(t\rightarrow\infty) = \tilde{\xi}(\epsilon=0)
\ee
and results in $p(t)= -\tanh\left(\beta\Delta\right)$ which is independent of the form of the memory function. The transient dynamics on the other hand would depend on the specific form of the memory functions. For example, the dynamics at short times would be dominated by the form of the term  $\frac{1}{\epsilon + \tilde{\mathcal{W}}_{12}(\epsilon) + \tilde{\mathcal{W}}_{21}(\epsilon)}$ in Eq.~(\ref{eq:plap}). 

\begin{figure}[!h]
\centering
\includegraphics[viewport = 10 0 300 220, width = 80 mm]{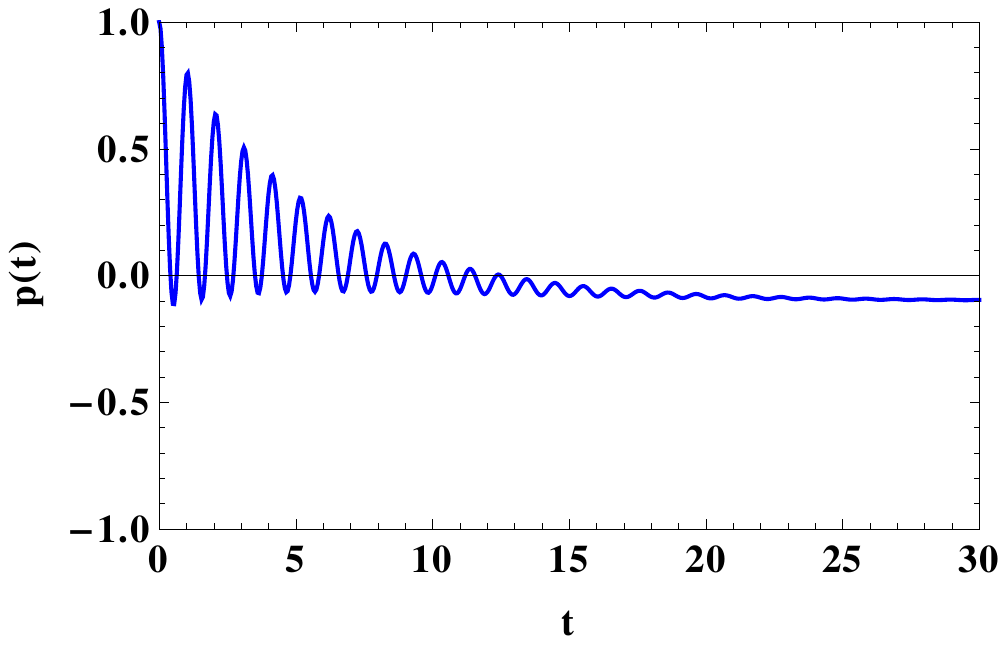}
\caption{Time evolution of $p$ resulting from the Lorentzian spectral function. The different parameters are $\beta = 0.1$, $\alpha=0.2$ and $\Delta = 1$.}\label{fig:lorentzyz}
\end{figure}

To further exemplify the connection provided by the GME between loss of coherence and decay to detailed balance let us consider a representative spectral function given by $\mathcal{Y}(z) = \frac{1}{1+ e^{-\beta z}} Y_{s}(z) $, which when substituted in Eqs.~(\ref{eq:ratesintform}) gives for the memories, 
\be\label{eq:memorylorentz}
\mathcal{W} = \frac{1}{2}\int_{-\infty}^{\infty}dz \left(1+ \tanh(\beta z/2)\right)Y_{s}(z)\cos\left[(z\pm2\Delta)t\right].
\ee
To proceed further, we need to assume some form for $Y_{s}(z)$. Any symmetric function $Y_{s}(z)$ would satisfy $\mathcal{Y}(-z) = \mathcal{Y}(z)e^{-\beta z}$ and would therefore be appropriate. As an example we consider a Lorentzian function for $Y_{s}(z)$  $(Y_{s}(z) = 1/\left(1 + \frac{z}{\alpha}\right)^{2})$. To keep the analysis analytically tractable we assume $\beta<<1$, and obtain the time evolution equation for $p$ from Eq.~(\ref{eq:2sitegme}) 
\ben\nonumber\label{eq:2sitegmelorentz}
\frac{dp}{dt} &+& \pi\alpha\int_{0}^{t}ds e^{-\alpha(t-s)}\cos\left(2\Delta(t-s)\right)p(s)\\&=&-\frac{\pi\alpha^{2}\beta}{2}\int_{0}^{t}ds~e^{-\alpha s}\sin(2\Delta s)
\een
In Fig.~\ref{fig:lorentzyz} we show both the coherent energy transfer for short times with decay of oscillations and finally the probability difference settling to a constant value 
obtained as solution of Eq.~(\ref{eq:2sitegmelorentz}). 

The steady state value of the probability difference at long times from Eq.~(\ref{eq:lontimelap}) is 
\be\label{eq:ltimelap}
\tilde{\xi}(\epsilon=0)= -\beta\Delta,
\ee
the first order term obtained in the expansion of $\tanh(\beta\Delta)$ for $\beta<<1$. The short time behavior is dominated by the term $\frac{1}{\epsilon + \tilde{\mathcal{W}}_{12}(\epsilon) + \tilde{\mathcal{W}}_{21}(\epsilon)}$ in Eq.~(\ref{eq:plap}) and for the memory under consideration is,  
\be\label{eq:pstime_lap}
\tilde{p}(\epsilon)= \frac{4\Delta^{2} + \epsilon^{2}\left(1 + \frac{\alpha}{\epsilon}\right)^{2}}{\epsilon\left( 4\Delta^{2} + \epsilon^{2}\left(1 + \frac{\alpha}{\epsilon}\right)^{2} + \pi\alpha\left(1 + \frac{\alpha}{\epsilon}\right)\right)},
\ee
which shows both oscillations of $p$ and its decay at short times. This loss of coherence at short times is a consequence of the coupling to the reservoir. For small values of $\alpha$ $(\alpha/\epsilon\sim 0)$ the oscillation seen at short times in Fig.~\ref{fig:lorentzyz} can be further approximated by  
\be\label{eq:pstime_lorentz}
p(t\rightarrow 0) = \frac{\pi\alpha\cos\left(\sqrt{4\Delta^{2} + \pi\alpha}~t\right) + 4\Delta^{2}}{4\Delta^{2} + \pi\alpha}.
\ee 
The observed dynamics of the probability difference can be therefore looked upon as the resultant of the interplay between three different effects. The first is coherent transfer with oscillations which occur between the maximum value of 1 and a minimum value, which is not -1 as in the degenerate case but  $\frac{4 \Delta^{2} - \pi\alpha}{4 \Delta^{2} + \pi/\alpha}$; this difference arises from the nondegeneracy (finite value of $\Delta$). The second is the decay of the coherence at short times as a result of coupling to the reservoir. The third is the ultimate decay to a steady state value consistent with the detailed balance condition, which, while it is also a result of the coupling to the reservoir, stems from a source different from that leading to the destruction of phase coherences.

\section{Application to nondegenerate dimer}\label{sec:nddimer}
As a more realistic example we now consider the standard model of a two-site nondegenerate dimer in interaction with bosons and apply the formalism developed in this paper to it. The Hamiltonian for this model is given by
\ben\nonumber\label{eq:hamelph}
H =& \hbar\omega\left(b^{\dg}b +\frac{1}{2}\right) + V\left(a_{1}^{\dg}a_{2} + a_{2}^{\dg}a_{1}\right) + \Delta\left(a_{1}^{\dg}a_{1} - a_{2}^{\dg}a_{2}\right) &\\ & + g\hbar\omega\left(b+b^{\dg}\right)\left(a_{1}^{\dg}a_{1} - a_{2}^{\dg}a_{2}\right)&
\een
\noindent
and consists of an electron interacting with phonons and tunneling between the two sites. In Eq.~(\ref{eq:hamelph}) $b$ and $b^{\dg}$ are respectively the annihilation and creation operator for the phonons, with a frequency $\omega$,  $a_{i}$ and $a_{i}^{\dg}$ are the particle's annihilation and creation operator at site i=$1,2$, $V$ is the tunneling matrix element between sites $1$ and $2$, $g$ is the electron-phonon coupling constant and $\Delta$ is the  the difference in energies at sites 1 and 2 which measures the nondegeneracy. Applying the unitary transformation $e^{S}He^{-S}$ with $S = g\left(b-b^{\dg}\right)\left(a_{1}^{\dg}a_{1} - a_{2}^{\dg}a_{2}\right)$ on Eq.~(\ref{eq:hamelph}) the polaron Hamiltonian given by
\ben\nonumber\label{eq:hampol}
H = &\Delta\left(A_{1}^{\dg}A_{1} - A_{2}^{\dg}A_{2}\right) + \hbar\omega\left(B^{\dg}B + \frac{1}{2}\right) &\\&+ V\left(A_{1}^{\dg}A_{2}e^{-2g\left(B -B^{\dg}\right)} + A_{2}^{\dg}A_{1}e^{2g\left(B -B^{\dg}\right)}\right)&
\een
\noindent
is obtained. In this transformed Hamiltonian the operators for the polaron and phonons operators are $B = b + g\left(a_{1}^{\dg}a_{1} - a_{2}^{\dg} a_{2}\right)$ and $A = ae^{g\left(b^{\dg} - b\right)}$ respectively. The transformed Hamiltonian of Eq.~(\ref{eq:hampol}), except for the last term, is diagonal in the combined basis of the polaron and phonons and hence is of the form $H = H_{0} + \mathcal{V}$ with $\mathcal{V} = V\left(A_{1}^{\dg}A_{2}e^{-2g\left(B -B^{\dg}\right)} + A_{2}^{\dg}A_{1}e^{2g\left(B -B^{\dg}\right)}\right)$ representing the off-diagonal contribution to the Hamiltonian responsible for the motion of the polaron between the two sites. The memory function obtained from the polaron Hamiltonian can be computed\cite{Rahman1974} (see also Eq.~(28) of ref. \cite{Kenkre2003}) and is given by
\be\label{eq:memorypol}
\mathcal{W}(t) = 2 e^{-4 g^{2}\left(1- \cos(\omega t)\right)\coth\left(\frac{\beta\omega}{2}\right)}\cos\left(4g^{2}\sin\omega t \pm 2 \Delta t \right).
\ee
The spectral function $\mathcal{Y}(z)$ of equations (\ref{eq:ratesintform}), (\ref{eq:wnmmodified}) and (\ref{eq:memorylorentz}) can be written in terms of a modified Bessel function, which is excellently approximated by a Gaussian for large temperatures and a Poisson distribution for small temperatures (see appendix for details), and is shown in Fig.~\ref{fig:yz} for two different temperatures values $\beta\omega= 0.1$ (left) and $\beta\omega= 20$ (right).
\begin{figure}[!h]
\centering
\includegraphics[viewport = 10 0 250 240, width = 38 mm]{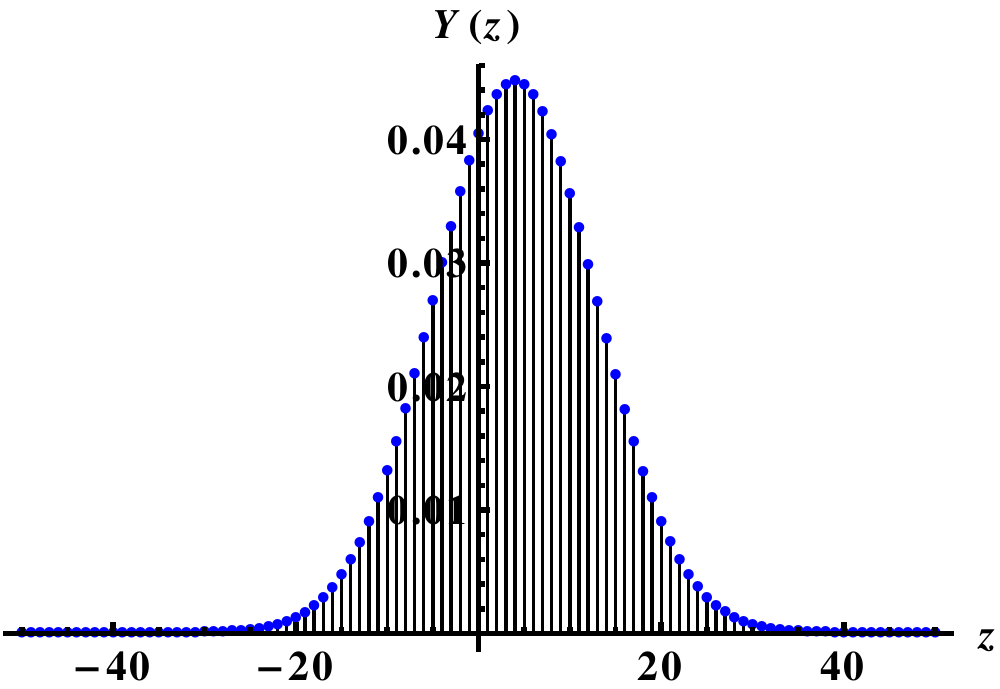}\hspace{0.25in}
\includegraphics[viewport = 10 0 250 240, width = 38 mm]{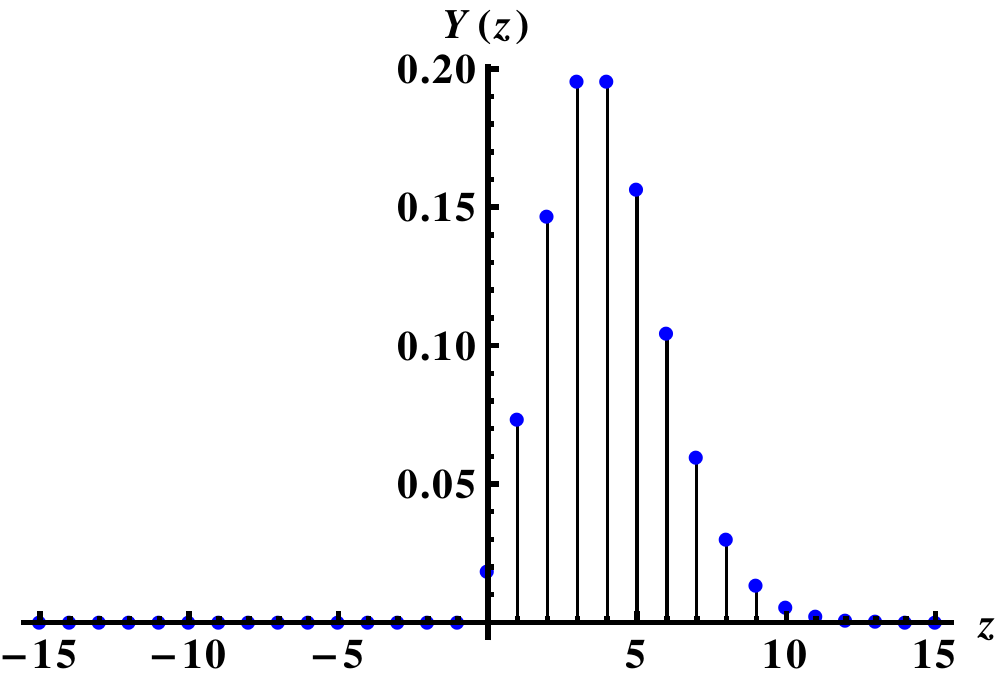}
\caption{Spectral function $\mathcal{Y}(z)$ in the high (left) and low (right) temperature limit normalized to $V^{2}$. The parameter values are $g=1$ and $\beta\omega= 0.1$ (left) and $\beta\omega= 20$ (right).}\label{fig:yz}
\end{figure}
The probability difference can be calculated by substituting the memory functions of Eq.~(\ref{eq:memorypol}) in Eq.~(\ref{eq:2sitegme}). Given our current interest which is in the general trends of the behavior,  we restrict the analysis to the case of  large temperatures $(\beta\omega<<1)$, in which case the spectral function can be expressed as a Gaussian (see Eq.~(\ref{eq:yzht}))
\be\label{eq:yzgauss}
\mathcal{Y}(z) = \frac{V^{2}}{2g \sqrt{\frac{\pi}{\beta\omega}}} e^{-\frac{\left(z - 4 g^{2}\right)^{2}}{16g^{2}/\beta\omega}}.
\ee
In Eq.~(\ref{eq:yzgauss}) we have assumed the density of states to be uniformly distributed, which would be applicable, for instance, in the case of acoustic phonons. From Eqs.~(\ref{eq:ratesintform}) the memory functions are  
\ben\label{eq:memgauss}
\mathcal{W}&=& 2 V^{2}e^{-\frac{4 g^{2}\omega t^{2}}{\beta}}\cos\left[(4 g^{2}\omega \pm 2\Delta)t\right].
\een
In Fig.~\ref{fig:longtime} we show the evolution of the probability difference for the memories given by Eq.~(\ref{eq:memgauss}). The parameters have been specifically chosen to clearly demonstrate the markedly different behaviors at short and long times. At short times the probability difference shows decaying oscillations around a fixed value (dashed line in Fig.~\ref{fig:longtime}). In this limit the probability is described by 
\be\label{eq:stark}
\frac{dp}{dt} = - \int_{0}^{t} ds \cos\left(\Delta(t-s)\right)p(s),
\ee
which is independent of the presence of the bath and whose solution, 
\be\label{eq:solstark}
p(t) = \frac{\Delta^{2}}{1 + \Delta^{2}} + \frac{1}{1 + \Delta^{2}}\cos\left( t\sqrt{1 + \Delta^{2}}\right),
\ee
shows coherent transfer, with $p$ oscillating around a displaced equilibrium value $\langle p\rangle = \Delta^{2}/\left(1 + \Delta^{2}\right)$. Coupling to the phonons leads to the loss of coherence and consequently, the decay of $p$ to the steady state (dashed-dotted line in Fig.~\ref{fig:longtime}) detailed balance value $(-\tanh(\beta\Delta))$. In addition, the long time taken by $p$ to reach equilibrium is due to the small value of the coupling to the phonons. The memory functions in Eqs.~(\ref{eq:ratesintform}) can be expressed as cosine and sine Fourier transforms of the spectral function. The Gaussian spectral function in the limit of vanishing coupling to the phonons tends to a  delta function, whose Fourier transform results in long-lived memory functions. 

\begin{figure}[!h]
\centering
\includegraphics[viewport = 10 0 300 200, width = 70 mm]{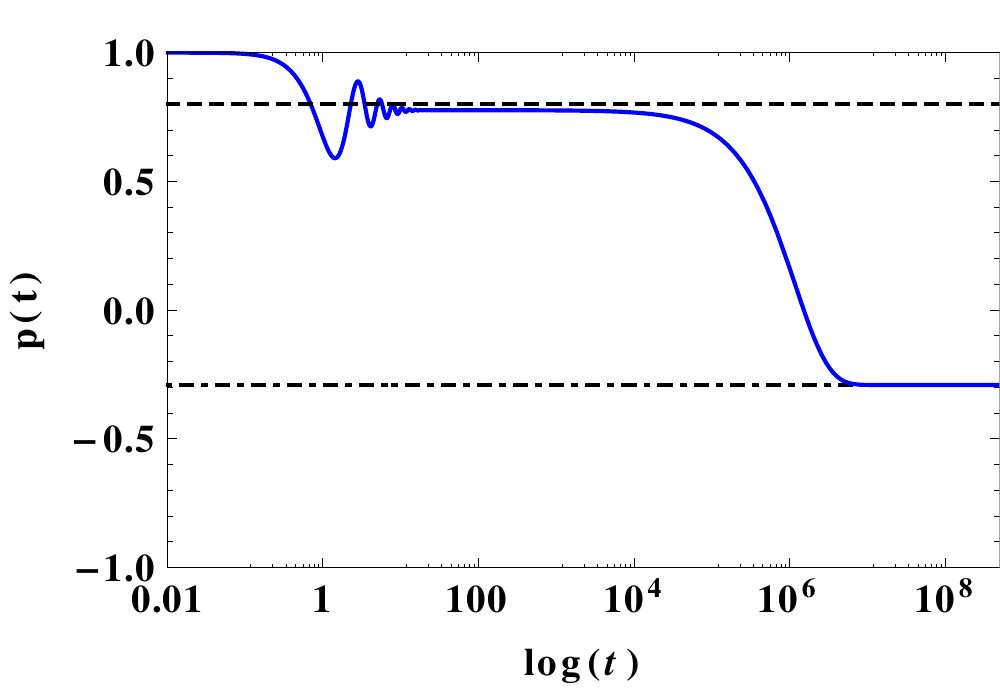}
\caption{Time evolution (solid line) of the probability difference as a function of dimensionless time $2Vt$ for Gaussian spectral function with uniform density of states showing two different regimes in which the short time behavior is dominated by the dc stark value (dashed line) and the long time behavior settles to the value given by the detailed balance condition (dashed-dotted line). The parameter values are $g^{2}\omega/V=0.02$, $\beta V=0.15$, $\Delta/V = 2$ and the initial condition is $p(0)=1$.}\label{fig:longtime}
\end{figure}

\section{Conclusions}\label{sec:conclusion}
In this paper we have developed a formalism of generalized master equation theory, appropriate, for describing the quantum evolution of a system, which, in addition to interaction with a reservoir is also \emph{nondegenerate}. Within the weak coupling approximation the theory presented combines the two distinctly different effects of the presence of reservoir, namely, loss of coherence and decay of probabilities to the value given by the Boltzmann factors. The important input required in the calculation ot the evolution of the probabilities are the time dependent memories appearing in the GME. They are determined via Fourier transforms of the spectral function $\mathcal{Y}(z)$. Particulars of the bath such as the density of states and the interaction with the subsystem of interest are contained within this spectral function, thus providing a complete general description of the quantum system of interest. The central result of our present analysis can be appreciated visually in Fig. 4 as the \emph{coexistence}
 of probability oscillations in the first part (short time)  and of probability decay to values displaying the correct Boltzmann probability ratios consonant with the nondegeneracy in the second part (long time). The first is characteristic of nondegeneracy in a quantum system while the second represents ultimately the equal \emph{a priori} postulate of equilibrium statistical mechanics. Note that the time scale in Fig. 4 is logarithmic: the two phenomena described accurately by the theory we have presented apply to \emph{widely} different times.While the examples presented in this paper are limited to two-site systems, the theory is completely 
general and is applicable to larger quantum systems\cite{ournote}. The transparent way in which the theory has been developed should allow for its application in the study of a large number of different systems.

\begin{appendix}
\section{High and Low temperature limits of the memory function}
The memory function of Eq.~(\ref{eq:memorypol}) can also be written as
\be
\mathcal{W}_{12}(t) = 2 V^{2} \exp\left[-4g^{2}\left((2\bar{n}+1) -(\bar{n}+1)e^{i\omega t} -\bar{n}e^{-i\omega t}\right)\right]
\ee
by using $\bar{n} = \frac{1}{e^{\beta\omega} - 1}$ $(\hbar=1)$ for the average number of phonons in thermal equilibrium and further by rearranging the terms as
\be
\mathcal{W}_{12}(t) = 2 V^{2} e^{-4g^{2}\left(2\bar{n}+1\right)}e^{4g^{2}\sqrt{\bar{n}(\bar{n}+1)}\left[\sqrt{\frac{\bar{n}+1}{\bar{n}}}e^{i\omega t} + \sqrt{\frac{\bar{n}}{\bar{n} +1}}e^{-i\omega t} \right]}
\ee
and using the relation $e^{\frac{x}{2}\left(t + \frac{1}{t}\right)} = \sum_{l=-\infty}^{\infty} I_{l}(x) t^{l}$ can be expressed in terms of the modified Bessel function
\be
\mathcal{W}_{12} = 2 V^{2} e^{-4g^{2}\left(2\bar{n}+1\right)} \sum_{z=-\infty}^{\infty} e^{i\omega z \left(t - \frac{i\beta}{2}\right)} I_{z}\left(8 g^{2}\sqrt{\bar{n}(\bar{n} + 1)}\right)
\ee
Comparing this with Eq.~(\ref{eq:ratesintform}) it is straightforward to see that the function $\mathcal{Y}(z)$ is given by
\be
\mathcal{Y}(z) = 2 V^{2} \rho(z) e^{-4g^{2}\left(2\bar{n}+1\right)} e^{-\frac{\beta\omega z}{2}} I_{z}\left(8 g^{2}\sqrt{\bar{n}(\bar{n} + 1)}\right)
\ee
which by using the expression for $\bar{n}$ can be written as
\be\label{eq:yz}
\mathcal{Y}(z) = 2 V^{2} \rho(z) e^{-4g^{2}\left(\frac{1 + k}{1-k}\right)} k^{-\frac{z}{2}} I_{z}\left(\frac{8 g^{2}\sqrt{k}}{1-k}\right)
\ee
where $k=e^{-\beta\omega}$. In the high temperature limit $T\rightarrow\infty$, $\beta\rightarrow 1$ for which $k\rightarrow1$. The large temperature expansion of Eq.~(\ref{eq:yz}) can be done by first realizing that when the argument of a modified Bessel function is much larger than its order than it can be approximated extremely well by a Gaussian, or $\mathcal{I}_{z}(x) \approx I_{0}(x)\exp(-z^{2}/2x)$. With this approximation, and using the asymptotic expansion for modified Bessel function in the case when the argument is much larger than the order and given by $I_{\alpha}(x)\approx \frac{e^{x}}{\sqrt{2\pi x}}$ we obtain for $\mathcal{Y}(z)$
\be
\mathcal{Y}(z)\approx \rho(z)\frac{V^{2}}  {2\sqrt{\pi} g \left(\frac{\sqrt{k}}{1-k}\right)^{1/2}} k^{-z/2} e^{-\frac{z^{2}}{16 g^{2}}\left(\frac{1-k}{\sqrt{k}}\right)}e^{-4 g^{2}\left(\frac{(1 - \sqrt{k})^{2}}{1-k}\right)}
\ee
which reduces to a Gaussian
\be\label{eq:yzht}
\mathcal{Y}(z) = \rho(z)\frac{V^{2}}{2g \sqrt{\frac{\pi}{\beta\omega}}} e^{-\frac{\left(z - 4 g^{2}\right)^{2}}{16g^{2}/\beta\omega}}
\ee
in the limit $k\rightarrow 1$. In the low temperature limit $(T\rightarrow 0$, $\beta\rightarrow\infty$ and hence $k\rightarrow 0)$ and using the small argument expansion of the modified Bessel function
\be
I_{\alpha}(x)\rightarrow \frac{1}{\Gamma(\alpha+1)}\left(\frac{x}{2}\right)^{\alpha}~~~0<x<<\sqrt{\alpha+1},
\ee
\noindent
in Eq.~(\ref{eq:yz}) we obtain a Poisson distribution for the function $\mathcal{Y}(z)$ at low temperatures
\be
\lim_{k\rightarrow 0}\mathcal{Y}(z) \approx V^{2}\rho(z)\left(\frac{e^{-4g^{2}}(4g^{2})^z}{z!}\right)\cdot
\ee
\end{appendix}


\begin{thebibliography}{}
\bibitem{history1} R.~S.~Knox, {\it  Excitation energy transfer and migration: Theoretical considerations in Bioenergetics of Photosynthesis} (Academic Press, New York 1975).
\bibitem{history2} R.~V.~Grondelle, J.~Amesz, {\it Excitation energy transfer in photosynthetic systems in Light Emission by Plants and Bacteria} (Academic Press, New York 1986).
\bibitem{history3} G.~Govindjee and R.~Govindjee, Scientific American {\bf 231}, 68 (1974).
\bibitem{history4} R.~K.~Clayton, J.~Theor. Biol. {\bf 14}, 173 (1967).
\bibitem{history5} R.~K.~Clayton, {\it Photosynthesis Physical Mechanisms and Chemical Patterns}, Cambridge University Press (1980).
\bibitem{Scholes2010}E.Collini, C.Y. Wong, K.E. Wilk, P.M.G. Curmi, P. Brumer and G.D. Scholes, Nature 463, 644 (2010).
\bibitem{Fleming2009} A.~Ishizaki, G.~R.~Fleming, Proc. Natl. Acad. Sci., {\bf 106}, 17255 (2009).
\bibitem{Forster1948} T.~F\"{o}rster, Ann. Phys. {\bf 437}, 55 (1948).
\bibitem{Kenkre1974} V.~M.~Kenkre, R.~S.~Knox, Phys.~Rev.~B {\bf 9}, 5279 (1974).
\bibitem{Silbey1971} M.~Grover, R.~Silbey, J.~Chem.~Phys. {\bf 54}, 4843 (1971).
\bibitem{Silbey1975} I.~I.~Abram, R.~Silbey, J.~Chem.~Phys. {\bf 63}, 2317 (1975).
\bibitem{Rackovsky1973} S.~Rackovsky, R.~Silbey, Mol. Phys. {\bf 25}, 61 (1973).
\bibitem{Zwanzig1961} R.~Zwanzig, {\it Statistical Mechanics of Irreversibility in lectures in Theoretical Physics}, Vol.~3 (Interscience Publisher Newyork, 1961).
\bibitem{Zwanzig2001} R.~Zwanzig, {\it Nonequilibrium Statistical Mechanics}, Oxford University Press (2001).
\bibitem{Kenkrer1} V.~M.~Kenkre, Phys.~Rev.~B {\bf 11}, 1741 (1975).
\bibitem{Kenkre1975} V.~M.~Kenkre, Phys.~Rev.~B {\bf 12}, 2150 (1975).
\bibitem{Kenkre1977} V.~M.~Kenkre, {\it The generalized master equation and it applications in stastical mechanics and statistical methods in theory and application} (New York, Plenum, 1977)
\bibitem{Rahman1974} V.~M.~Kenkre,T.~S.~Rahman, Phys.~Lett {\bf 50 A}, 170 (1974).
\bibitem{Kenkre2003} V.M. Kenkre, in {\it Modern Challenges in Statistical Mechanics: Patterns, Noise, and the Interplay of Nonlinearity and Complexity}, edited by V.M. Kenkre and K. Lindenberg, AIP Conf. Proc. No. 658
(AIP, Melville, NY, 2003).
\bibitem{ournote} We have not
merely repeated the standard textbook-type steps in a Zwanzig form derivation in section II as is sometimes
done. The detail laid out there and comments made are important as they precisely explain how the
detailed balance nature of rates occurs even after a weak-coupling approximation of the expressions.
\end{thebibliography}
\end{document}